\documentclass[a4paper,12pt]{article}

\usepackage{amsmath}

\def \gsim
{\raisebox{-3pt}{$\>\stackrel{>}{\scriptstyle\sim}\>$}}

\def\beq{\begin{equation}}
\def\be{\begin{equation}}
\def\eeq{\end{equation}}
\def\ee{\end{equation}}
\def\bea{\begin{eqnarray}}
\def\eea{\end{eqnarray}}

\def\bq{\begin{quote}}
\def\eq{\end{quote}}

\def\gappeq{\mathrel{\rlap {\raise.5ex\hbox{$>$}}
{\lower.5ex\hbox{$\sim$}}}}
\def\lappeq{\mathrel{\rlap{\raise.5ex\hbox{$<$}}
{\lower.5ex\hbox{$\sim$}}}}
\def\Toprel#1\over#2{\mathrel{\mathop{#2}\limits^{#1}}}

\allowdisplaybreaks

\begin{document}

\begin{titlepage}
\vspace*{-1.2cm}
\begin{flushright}
\small
ROME1/1466-08\\
April 2008
\end{flushright}
\vspace{1.cm}
{\Large
\begin{center}
{\bf  Comment on Resummation of Mass Distribution
\\ for Jets Initiated by Massive Quarks}
\end{center}
}
\vspace{.8cm}

\begin{center}
{ 
\bf{U.\ Aglietti~$^{a}$,~~L.\ Di~Giustino~$^{b}$,~~G.\ Ferrera~$^{c}$,
~~L.\ Trentadue~$^{b}$}}
\\[9mm]
\small {$^{a}$\textit{Dipartimento di Fisica, Universit\`a di Roma ``La Sapienza''
 and\\
INFN Sezione di Roma, Roma, Italy }}
\\[2mm]
{$^{b}$\textit{Dipartimento di Fisica, Universit\`a di Parma
 and\\
INFN,~ Gruppo Collegato di Parma, Parma, Italy}}
\\[2mm]
{$^{c}$\textit{Dipartimento di Fisica, Universit\`a di Firenze and\\
INFN Sezione di Firenze, Firenze, Italy}}\\
[10pt] \vspace{1.5cm}
\begin{abstract}

We compute in the heavy quark effective theory the soft coefficient
$D_2$ entering the resummation of next-to-next-to-leading threshold 
logarithms for jets initiated by a quark with a small mass compared
to the hard scale of the process.
We find complete agreement with a previous computation in full QCD.
Contrary to our previous guess, this coefficient turns 
out to be different from that one entering heavy flavor decay or 
heavy flavor fragmentation.

\end{abstract}
~\\
\end{center}
\end{titlepage}

\setcounter{footnote}{0} \setcounter{page}{2} \setcounter{section}{0}
\newpage

\noindent
In~\cite{Aglietti:2006wh} we have considered the resummation of threshold
logarithms for jets initiated by a quark with a small mass compared to
the hard scale of the process.
The main result is the following factorization formula, in $N$-moment space,
for the probability that a quark with virtuality $Q$ and mass $m \ll Q$ 
fragments into a jet of mass $m_X$:
\beq 
\label{univ} 
J_N(Q^2; \, m^2)
\, = \, J_N(Q^2) ~ \delta_N(Q^2; \, m^2) \, , 
\eeq
where $J_N(Q^2)$ is the standard massless fragmentation function,
whose resummed expression reads \cite{catanitrentadue,Catani:1990rr,catcac}:
\beq
J_N(Q^2) \, = \, \exp\!
\int_0^1 d y \, \frac{ (1-y)^{N-1} - 1 }{y}
\Bigg\{
\, \int_{ Q^2 y^2 }^{ Q^2 y } \frac{dk_{\perp}^2}{k_{\perp}^2} 
\,A\left[\alpha_S\left(k_{\perp}^2\right)\right]
\, + \, B \left[\alpha_S\left( Q^2 y \right)\right] \,
\Bigg\} \, ,
\eeq
and $\delta_N(Q^2; \, m^2)$ is the mass correction factor, 
whose resummed expression reads for $N \gsim 1/r$:
\beq
\label{maineq}
\hskip -0.1truecm
\delta_N(Q^2; m^2) = \exp\!
\int_0^1 \!\!\!dy\, \frac{ (1-y)^{r\,(N-1)}-1}{y}
\Bigg\{\!
 -\!\! \int_{ m^2 y^2 }^{ m^2 y } 
\frac{dk_{\perp}^2}{k_{\perp}^2} A\big[\alpha_S(k_{\perp}^2)\big]
 - B\big[\alpha_S( m^2 y )\big]
+ D\big[\alpha_S( m^2 y^2 )\big]
\!\!\Bigg\} .
\eeq
The functions $A(\alpha_S)$, $B(\alpha_S)$ and $D(\alpha_S)$, entering the
above resummation formulae, have a perturbative expansion in powers of $\alpha_S$:
\beq
A(\alpha_S) \, = \, \sum_{n=1}^{\infty} A_n \, \alpha_S^n \, ; \qquad
B(\alpha_S) = \sum_{n=1}^{\infty} B_n\,\alpha_S^n \,;\qquad 
D(\alpha_S) = \sum_{n=1}^{\infty} D_n\,\alpha_S^n \, ,
\eeq
and we have defined\,\footnote{
The case of a non-relativistic motion  of a massive color charge,
$m \, \approx \, \mathcal{O}(Q)$, which does not give rise to a jet, has been
analyzed in \cite{Aglietti:2007bp}.
}:
\beq
\label{eqdefy}
y \, \equiv \, \frac{m_X^2 - m^2}{Q^2 - m^2} \, ;
\qquad
r \, \equiv \, \frac{m^2}{Q^2} \, \cong \, \frac{m^2}{Q^2 - m^2} \, \ll \, 1 \, .
\eeq
Since the functions $A(\alpha_S)$ and $B(\alpha_S)$ are related 
to small-angle emission only, they represent universal intra-jet 
properties, as confirmed by explicit higher-order computations; 
at present, the first three coefficients of both these functions are analytically 
known \cite{eroi}, allowing for next-to-next-to-leading logarithmic accuracy. 
On the contrary, the function $D(\alpha_S)$, being related to soft emission 
at large angle with respect to the quark, is in general a process-dependent 
inter-jet quantity.
In the framework of fragmentation functions, soft radiation
not collinearly enhanced is described by the function 
$D^{(f)}(\alpha_S)$ \cite{catcac}, while in heavy flavor decay
it is described by the function $D^{(h)}(\alpha_S)$, which has been shown to coincide
with the former one to all orders in $\alpha_S$ in \cite{gardi}.
In~\cite{Aglietti:2006wh} we explicitly computed the coefficient $D_1$,
which turned out to be equal to $D_1^{(f,h)}$.
On the basis of a physical argument, we then conjectured that the equality
could extend to higher orders, i.e. that $D(\alpha_S) = D^{(f,h)}(\alpha_S)$.
After the publication of~\cite{Aglietti:2006wh}, the coefficient $D_2$
was computed in full QCD in \cite{Mitov:2006xs} by using explicit results 
on massive two-loop vertex functions \cite{Bernreuther:2004ih}.
This computation showed explicitly that $D_2 \ne D_2^{(f,h)}$, invalidating 
our previous guess $D(\alpha) \, = \, D^{(f,h)}(\alpha)$.
In this note we present an independent recomputation of $D_2$ in 
the heavy quark effective theory, based on an expansion 
of the two-loop cusp anomalous dimension $\Gamma_{\rm cusp}$ \cite{Korchemsky:1987wg}.  
As in~\cite{Aglietti:2006wh}, let us consider the case of the $b \, \to \, s \, \gamma$ decay,
in the rest frame of the $b$-quark. The hard scale of the process 
is given by the beauty mass, $Q = m_b$, and the mass correction parameter
by $r = m_s^2/m_b^2 \ll 1$.
The Minkowskian cusp angle then reads
\beq
\gamma = \mbox{arccosh} \, \left( v_b \cdot v_s \right) \, = \, \frac{1}{2} \log \frac{1}{r} \, ,
\eeq
where $v_b$ and $v_s$ are the $b$ and $s$ quark 4-velocities ($v_b^2 = v_s^2 = 1$).
By expanding for $r \ll 1$ (two times) the cusp anomalous dimension  
evaluated in~\cite{Korchemsky:1987wg}, one obtains:
\begin{eqnarray}
2\, \Gamma_{\rm cusp}(r; \, \alpha_S) \, \cong \, 
\left[
A_1 \alpha_S + A_2 \alpha_S^2 
\right] \log \frac{1}{r}
\, + \,  
\left[
D_1 + D_1^{(h)} 
\right] 
\alpha_S
\, + \, 
\left[
D_2 + D_2^{(h)} 
\right] 
\alpha_S^2,
\end{eqnarray}
where
\begin{eqnarray}
D_1 \, = \, D_1^{(h)} &=& - \, \frac{C_F}{\pi} \, ;
\\
D_2 \, + \, D_2^{(h)} &=& \frac{C_F}{\pi^2} 
\left[ 
C_A
\left(
z(2) - z(3)  - \frac{49}{18}  
\right)
- \frac{5}{18} n_f
\right] \, . 
\end{eqnarray}
Let us stress that we have two different $D$-like contributions to the cusp anomalous
dimension: $D^{(h)}(\alpha)$, related to soft emission off the decaying $b$-quark,
and $D(\alpha)$, related to the soft emission off the massive $s$-quark originating 
the jet in the final state\,\footnote{
There is no $B$-like contribution to the cusp anomalous dimension because the heavy quark 
effective theory only describes soft interactions.
}.
Fermionic contributions, not evaluated in~\cite{Korchemsky:1987wg}, have been added by
noting that they are of the form $\alpha_S^2 \, n_f ( \log 1/r - 2 )$
and by including in the coefficient $A_2$ above also the known $\mathcal{O}(n_f)$ contribution.
By assuming that the eikonal result above can be directly related to full QCD 
and by subtracting from the sum above the standard $D_2^{(h)}$ for 
heavy flavor decay,
\begin{eqnarray}
D_2^{(h)} \, = \,  
\frac{C_F}{\pi^2} 
\left[
C_A 
\left(
- \frac{9}{4} z(3)
+ \frac{1}{2} z(2)
+ \frac{55}{108} 
\right)
+ \frac{n_f}{54}
\right] 
\simeq  - 0.556416 + 0.002502 \,n_f ,
\end{eqnarray}
we obtain for the coefficient entering the jet-correction factor $\delta_N(Q^2,m^2)$:
\begin{eqnarray}
D_2 \, =  \,
\frac{C_F}{\pi^2} 
\left[
C_A 
\left(
\frac{5}{4} z(3)
+ \frac{1}{2} z(2)
- \frac{349}{108}
\right)
+ \frac{29}{54}\,n_f
\right] 
\, \simeq - 0.367368 + 0.072551\,n_f .
\end{eqnarray}
The above expression for $D_2$ is in complete 
agreement with that one obtained in full
QCD in Eq.~(68) of \cite{Mitov:2006xs}, 
both in the Abelian and non-Abelian pieces
(after a trivial difference in normalization is taken into account).
As a consequence, the expressions for the NNLL coefficients $H_{21}$ and $H_{32}$ 
given in Eqs.~(141) and (144)  of \cite{Aglietti:2006wh} respectively are 
also incorrect and must be replaced by:
\begin{eqnarray}
H_{21} &=&
  \frac{ C_F^2}{2 \pi ^2} \left( - \frac{3}{16} + \frac{\pi ^2}{6} - 5 z(3)\right)
+\frac{C_F C_A}{4 \pi ^2} \left(5 z(3) +\frac{5 \pi ^2 }{18} - \frac{121}{72}\right) \, +
\nonumber
\\
&& + \, \frac{C_F n_f}{\pi^2} \left( \frac{5}{144} - \frac{\pi^2}{36} \right)    \, ;
\\
H_{32}&=& 
- \frac{C_F^3}{4 \pi ^3} \left( \frac{\pi ^4}{90} + z(3)\right)
+ \frac{ C_A C_F^2}{4 \pi ^3} \left( - \frac{11}{32} - \frac{767 \pi ^2}{432} + \frac{\pi ^4}{18} - 22 z(3)\right) +
\nonumber\\
&& + \, \frac{C_A^2 C_F}{4 \pi ^3} \left(\frac{44539}{2592}-\frac{145 \pi^2}{216}
+\frac{11 \pi ^4}{360}+\frac{11 z(3)}{12}\right) 
+ \frac{n_f^2 C_F}{4 \pi ^3} \left(\frac{205}{648}+\frac{\pi ^2}{54}\right)
\, + \nonumber\\
&&
\, + \, \frac{n_f C_F^2}{4 \pi^3} \left(-\frac{67}{48} + \frac{61 \pi^2}{216}+5 z(3)\right) 
+ \frac{ C_A n_f C_F }{4 \pi^3} \left(-\frac{6745}{1296}-\frac{7 z(3)}{6}\right) \, .
\end{eqnarray}

\vskip 1truecm
\centerline{\bf Acknowledgments}
\vskip 0.1truecm

One of us (U.~A.) would like to thank E.~Remiddi for discussions.


\newpage
\begin{thebibliography}{99}

\bibitem{Aglietti:2006wh}
U.~Aglietti, L.~Di Giustino, G.~Ferrera and L.~Trentadue,
{\itshape  ``Resummed mass distribution for jets initiated by massive quarks,''}
Phys.\ Lett.\  B {\bf 651}, 275 (2007) [arXiv:hep-ph/0612073].

\bibitem{catanitrentadue}
S.~Catani and L.~Trentadue,
{\itshape ``Resummation of the QCD Perturbative Series for Hard Processes,''}
Nucl.\ Phys.\ B {\bf 327} (1989) 323.

\bibitem{Catani:1990rr}
S.~Catani, B.~R.~Webber and G.~Marchesini, 
{\itshape ``QCD coherent branching and semi-inclusive processes at large $x$,''}
Nucl.\ Phys.\ B {\bf 349} (1991) 635.

\bibitem{catcac}
M.~Cacciari and S.~Catani, 
{\itshape ``Soft-gluon resummation for the fragmentation of light and heavy quarks  at large $x$,''}
Nucl.\ Phys.\ B {\bf 617} (2001) 253 [arXiv:hep-ph/0107138].

\bibitem{Aglietti:2007bp}
U.~Aglietti, L.~Di Giustino, G.~Ferrera, A.~Renzaglia, G.~Ricciardi and 
L.~Trentadue,
{\itshape  ``Threshold Resummation in $B \to X_c \,\,l\,\, \nu_l$ Decays,''}
Phys.\ Lett.\  B {\bf 653} (2007) 38 [arXiv:0707.2010 [hep-ph]].

\bibitem{eroi}
S.~Moch and A.~Vogt,
{\itshape ``Higher-order soft corrections to lepton pair and Higgs boson  production,''}
Phys.\ Lett.\  B {\bf 631} (2005) 48 [arXiv:hep-ph/0508265].

\bibitem{gardi}
E.~Gardi, {\itshape ``On the quark distribution in an on-shell heavy quark and 
its all-order relations with the perturbative fragmentation function,'' }
JHEP {\bf 0502} (2005) 53 [arXiv:hep-ph/05010257].



\bibitem{Mitov:2006xs}
A.~Mitov and S.~Moch, {\itshape 
``The singular behavior of massive QCD amplitudes,''}
JHEP {\bf 0705} (2007) 001 [arXiv:hep-ph/0612149].

\bibitem{Bernreuther:2004ih}
  W.~Bernreuther, R.~Bonciani, T.~Gehrmann, R.~Heinesch, T.~Leineweber, P.~Mastrolia and E.~Remiddi,
 {\itshape
   ``Two-loop QCD corrections to the heavy quark form factors: The vector
  contributions,''}
  Nucl.\ Phys.\  B {\bf 706} (2005) 245
  [arXiv:hep-ph/0406046].


\bibitem{Korchemsky:1987wg}
G.~P.~Korchemsky and A.~V.~Radyushkin,
{\itshape ``Renormalization of the Wilson Loops Beyond the Leading Order,''}
Nucl.\ Phys.\  B {\bf 283} (1987) 342.

\end{thebibliography}
\end{document}